\documentclass[aps,prl,reprint,noeprint,superscriptaddress]{revtex4-1}
\usepackage{graphicx}
\usepackage{braket}
\usepackage{amsmath}
\usepackage{bm}
\usepackage{bbold}
\usepackage{url}
\usepackage{amssymb}
\usepackage[autostyle=true,english=american]{csquotes}
\usepackage{xcolor}

\begin{document}

\title{Coherent long-distance spin-qubit--transmon coupling}

\author{A. J. Landig}
\affiliation{Department of Physics, ETH Z\"urich, CH-8093 Z\"urich, Switzerland}
\author{J. V. Koski}
\affiliation{Department of Physics, ETH Z\"urich, CH-8093 Z\"urich, Switzerland}
\author{P. Scarlino}
\affiliation{Department of Physics, ETH Z\"urich, CH-8093 Z\"urich, Switzerland}
\author{C. M\"uller}
\affiliation{IBM Research Zurich, CH-8803 R\"uschlikon, Switzerland}
\author{J. C. Abadillo-Uriel}
\affiliation{Department of Physics, University of Wisconsin-Madison, Madison, WI 53706, United States}
\author{B. Kratochwil}
\affiliation{Department of Physics, ETH Z\"urich, CH-8093 Z\"urich, 
Switzerland}
\author{C. Reichl}
\affiliation{Department of Physics, ETH Z\"urich, CH-8093 Z\"urich, Switzerland}
\author{W. Wegscheider}
\affiliation{Department of Physics, ETH Z\"urich, CH-8093 Z\"urich, Switzerland} 
\author{S. N. Coppersmith}
\altaffiliation{current address: School of Physics, University of New South Wales, Sydney NSW 2052, Australia}
\affiliation{Department of Physics, University of Wisconsin-Madison, Madison, WI 53706, United States}
\author{Mark Friesen}
\affiliation{Department of Physics, University of Wisconsin-Madison, Madison, WI 53706, United States}
\author{A. Wallraff}
\affiliation{Department of Physics, ETH Z\"urich, CH-8093 Z\"urich, Switzerland} 
\author{T. Ihn}
\affiliation{Department of Physics, ETH Z\"urich, CH-8093 Z\"urich, Switzerland}
\author{K. Ensslin}
\affiliation{Department of Physics, ETH Z\"urich, CH-8093 Z\"urich, Switzerland}

\begin{abstract}
Spin qubits and superconducting qubits are among the promising candidates for a solid state quantum computer. For the implementation of a hybrid architecture which can profit from the advantages of either world, a coherent long-distance link is necessary that integrates and couples both qubit types on the same chip. We realize such a link with a frequency-tunable high impedance SQUID array resonator. The spin qubit is a resonant exchange qubit hosted in a GaAs triple quantum dot. It can be operated at zero magnetic field, allowing it to coexist with superconducting qubits on the same chip. We find a working point for the spin qubit, where the ratio between its coupling strength and decoherence rate is optimized. We observe coherent interaction between the resonant exchange qubit and a transmon qubit in both resonant and dispersive regimes, where the interaction is mediated either by real or virtual resonator photons.

\end{abstract}

\maketitle

\section{Introduction}
A future quantum processor will benefit from the advantages of different qubit implementations \cite{Acin2018}. Two prominent workhorses of solid state qubit implementations are spin- and superconducting qubits. While spin qubits have a high anharmonicity, a small footprint \cite{Hanson2007} and promise long coherence times \cite{Veldhorst2014,Yoneda2018,Russ2018}, superconducting qubits allow fast and high fidelity read-out and control \cite{OMalley2014,Walter2017}. To integrate both qubit systems on one scalable quantum device, a coherent long-distance link between the two is required. A technology to implement such a link is circuit quantum electrodynamics (cQED) \cite{Blais2004}, where microwave photons confined in a superconducting resonator couple coherently to the qubits. cQED was initially developed for superconducting qubits \cite{Wallraff2004}, where long-distance coupling \cite{Majer2007,Sillanpaa2007} enables two-qubit gate operations \cite{DiCarlo2009}. Recently, coherent qubit-photon coupling was demonstrated for spin qubits \cite{Mi2018,Samkharadze2018,Landig2018} in few electron quantum dots. However, coupling a spin qubit to another distant qubit has not yet been shown. One major challenge for an interface between spin and superconducting qubits is that spin qubits typically require large magnetic fields \cite{Loss1998,Petta2005}, to which superconductors are not resilient \cite{Luthi2018}. 

We overcome this challenge by using a spin qubit that relies on exchange interaction \cite{DiVincenzo2000}. This resonant exchange (RX) qubit \cite{Gaudreau2011,Medford2013,Medford2013a,Taylor2013,Russ2015} is formed by three electrons in a GaAs triple quantum dot (TQD). We implement the qubit at zero magnetic field without reducing its coherence compared to earlier measurements at finite magnetic field \cite{Landig2018}. The quantum link is realized with a frequency-tunable high impedance SQUID array resonator \cite{Stockklauser2017}, that couples the RX and the superconducting qubit coherently over a distance of a few hundred micrometers. The RX qubit coupling strength to the resonator and its decoherence rate are tunable electrically. We find that their ratio is comparable to previously reported values for spin qubits in Si \cite{Mi2018,Samkharadze2018}. We demonstrate coherent coupling between the two qubits first by resonant and then by virtual photons in the quantum link. Thereby we electrostatically tune the RX qubit to different regimes, where the qubit states have either a dominant spin or charge character. We also report that the SQUID array resonator can affect the qubit performance, which we suspect to be caused by charge noise introduced through the resonator.

\section{Sample and qubit characterization}
The design of our sample is illustrated schematically in Fig.~1(a). It is similar to Ref.~\onlinecite{Scarlino2018}, where the focus was on charge qubits. We use a superconducting qubit in the standard transmon configuration \cite{Koch2007,Houck2008}. It consists of an Al SQUID grounded on one side and connected in parallel to a large shunt capacitor. We tune the transition frequency $\nu_\mathrm{T}$ between the transmon ground $\ket{0_\mathrm{T}}$ and first excited state $\ket{1_\mathrm{T}}$ by changing the flux $\Phi_\mathrm{T}$ through the SQUID loop with an on-chip flux line. 

The transmon and the RX qubit are capacitively coupled to the same end of a SQUID array resonator, which we denote as coupling resonator in the following, with electric dipole coupling strengths $g_\mathrm{T}$ and $g_\mathrm{RX}$. The other end of the coupling resonator is connected to DC ground. It is fabricated as an array of Al SQUID loops \cite{Stockklauser2017}, which enables us to tune its resonance frequency $\nu_\mathrm{C}$ within a range of a few GHz with a magnetic flux $\Phi_\mathrm{C}$ produced by a coil mounted close to the sample. In addition, the resonator has a high characteristic impedance that enhances its coupling strength to both qubits. The transmon flux $\Phi_\mathrm{T}$ has a negligible effect on $\nu_\mathrm{C}$. 

The transmon is also capacitively coupled to a $50\,\mathrm{\Omega}$ $\lambda/2$ coplanar waveguide resonator with a coupling strength $g_\mathrm{R}/2\pi\simeq141\,\mathrm{MHz}$. Throughout this article, we refer to this resonator as the read-out resonator, because it allows us to independently probe the transmon without populating the coupling resonator with photons. The read-out resonator has a bare resonance frequency $\nu_\mathrm{R}=5.62\,\mathrm{GHz}$ and a total photon decay rate $\kappa_\mathrm{R}/2\pi=5.3\,\mathrm{MHz}$. As illustrated in Fig.~1(a), coupling and read-out resonators are measured in reflection mode by multiplexing a single probe tone at frequency $\nu_\mathrm{p}$. In addition, we can apply a drive tone at frequency $\nu_\mathrm{d}$ to both qubits via the resonators. For the experiments presented in this
work, the probe tone power is kept sufficiently low to ensure that the average number of photons in both resonators is less than one.

In Fig.~1(b) we characterize the transmon with two-tone spectroscopy. The first tone probes the read-out resonator on resonance ($\nu_\mathrm{p}=\nu_\mathrm{R}$), while the second tone is a drive at frequency $\nu_\mathrm{d}$ that is swept to probe the transmon resonance. Once $\nu_\mathrm{d}=\nu_\mathrm{T}$, the transmon is driven to a mixed state, which is observed as a change in the resonance frequency of the dispersively coupled read-out resonator. This frequency shift is detected with a standard heterodyne detection scheme \cite{Frey2012} as a change in the complex amplitude $\bm{A}=I+iQ$ of the signal reflected by the resonator. In Fig.~1(b) we observe a peak in $|\bm{A}-\bm{A}_0|$ centered at $\nu_\mathrm{d} = \nu_\mathrm{T}$. Here, $\bm{A}_0$ is the complex amplitude in the absence of the drive. From a fit of the transmon dispersion to the multi-level Jaynes-Cummings model and by including the position of higher excited states of the transmon probed by two photon transitions (not shown) \cite{Schuster2005,Schreier2008}, we obtain the maximum Josephson energy $E_\mathrm{J,max}=18.09\,\mathrm{GHz}$ and the transmon charging energy $E_\mathrm{c}=0.22\,\mathrm{GHz}$. 

At a distance of approximately $200\,\mathrm{\mu m}$ from the transmon, we form a TQD by locally depleting a two-dimensional electron gas in a GaAs/AlGaAs heterostructure with the Al top gate electrodes shown in Fig.~1(c). One of the electrodes directly extends to the coupling resonator to enable electric dipole interaction between TQD states and coupling resonator photons. Another electrode allows us to apply RF signals at frequency $\nu_\mathrm{dRX}$. We use a QPC charge detector to help tune the TQD to the three electron regime. The symmetric $(1,1,1)$ and the asymmetric $(2,0,1)$ and $(1,0,2)$ charge configurations are relevant for the RX qubit as they are lowest in energy. We sweep combinations of voltages on the TQD gate electrodes to set the energy of the asymmetric configurations equal and control the energy detuning $\Delta$ of the symmetric configuration with respect to the asymmetric ones [see Fig.~1(d)]. There are two spin states within $(1,1,1)$ that have $S = S_z = 1/2$ equal to the spin of two states with asymmetric charge configuration, which form a singlet in the doubly occupied dot. An equivalent set of states with $S = 1/2$, $S_z =-1/2$ exists. As the resonator response is identical for both sets of states, considering only one is sufficient (see Ref.~\onlinecite{Supplement} for a detailed discussion). This results in a total of four relevant states for the qubit \cite{Landig2018}. The tunnel coupling $t_\mathrm{l}$ ($t_\mathrm{r}$) between the left (right) quantum dot and the middle quantum dot hybridizes these states, which leads to the formation of the two RX qubit states $\ket{0_\mathrm{RX}}$ and $\ket{1_\mathrm{RX}}$. 
\begin{figure}[t!]
\includegraphics[bb=0 0 244 179,width=1\linewidth]{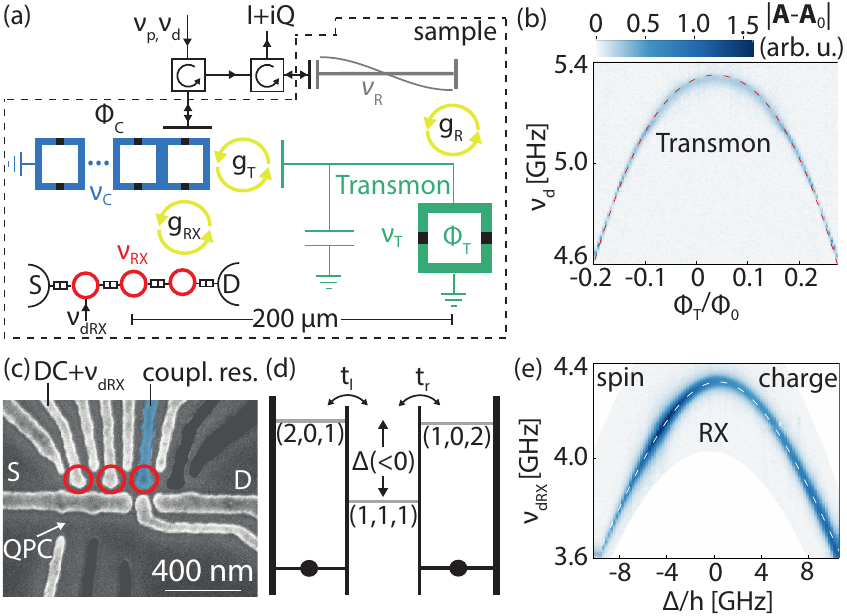}
\caption{\label{Fig1} Sample and qubit dispersions. (a) Schematic of sample and measurement scheme. The signals at frequencies $\nu_\mathrm{p}$ (probe) and $\nu_\mathrm{d}$ (drive) are routed with circulators as indicated by arrows. The reflected signal $I+iQ$ at $\nu_\mathrm{p}$ is measured. The sample (dashed line) contains four quantum systems with transition frequencies $\nu_{i}$: a coupling resonator that consists of an array of SQUID loops ($\nu_\mathrm{C}$, blue), an RX qubit ($\nu_\mathrm{RX}$, red), a transmon ($\nu_\mathrm{T}$, green) and a read-out resonator ($\nu_\mathrm{R}$, gray). Empty black double-squares indicate electron tunnel barriers separating the three quantum dots (red circles) as well as the source (S) and drain (D) electron reservoirs. A drive tone at frequency $\nu_\mathrm{dRX}$ can be applied to one of the dots. Filled black squares denote the Josephson junctions of SQUIDs. Yellow circles with arrows mark coupling between the quantum systems with coupling strengths $g_i$. $\Phi_\mathrm{C}$ and $\Phi_\mathrm{T}$ denote coupling resonator and transmon flux, respectively. (b) Two-tone spectroscopy of the transmon, with the RX qubit energetically far detuned. We plot the complex amplitude change $|\bm{A}-\bm{A}_0|$ (see main text) as a function of drive frequency $\nu_\mathrm{d}$ and $\Phi_T/\Phi_0$. The dashed line indicates the calculated $\nu_{T}$. (c) Scanning electron micrograph of the TQD and quantum point contact (QPC) region of the sample. Unused gate lines are grayed out. The gate line extending to the coupling resonator is highlighted in blue. (d) TQD energy level diagram indicating the tunnel couplings $t_\mathrm{l}$ and $t_\mathrm{r}$ and the electrochemical potentials, parametrized by $\Delta$, of the relevant RX qubit states ($N_\mathrm{l}$,$N_\mathrm{m}$,$N_\mathrm{r}$) with $N_\mathrm{l}$ electrons in the left, $N_\mathrm{m}$ electrons in the middle and $N_\mathrm{r}$ electrons in the right quantum dot. (e) Two-tone spectroscopy of the RX qubit, with the transmon energetically far detuned for $\nu_p\simeq\nu_\mathrm{C}=4.84\,\mathrm{GHz}$ as a function of $\Delta$ and $\nu_\mathrm{dRX}$. The dashed line shows the expected qubit energy as obtained from theory.}
\end{figure}
For $\Delta<0$, $\ket{0_\mathrm{RX}}$ and $\ket{1_\mathrm{RX}}$ have predominantly the $(1,1,1)$ charge configuration but different spin arrangement. Consequently, quantum information is predominantly encoded into the spin degree of freedom. With increasingly negative $\Delta$, the spin character of the qubit increases, which reduces the qubit dephasing due to charge noise. This comes at the cost of a reduced admixture of asymmetric charge states and therefore a decrease in the electric dipole coupling strength $g_\mathrm{RX}$. In contrast, for $\Delta>0$ the RX qubit states have dominantly asymmetric charge configurations $(2,0,1)$ and $(1,0,2)$. The qubit therefore has a dominant charge character, which increases, together with $g_\mathrm{RX}$, with increasing positive $\Delta$. 
Independent of $\Delta$, the RX qubit states have the same total spin and spin $z$-component such that they can directly be driven by electric fields \cite{Russ2017} and be operated in the absence of an applied external magnetic field. This is in contrast to other spin qubit implementations, which rely on engineered or intrinsic spin-orbit interaction \cite{Tokura2006,Pioro-Ladriere2008,Golovach2006,Nowack2007,Cottet2010,Viennot2015} for spin-charge coupling. 

Four similar RX qubit tunnel coupling configurations were used in this work as listed in Table \ref{QubitConfigsTable}. We use two-tone spectroscopy \cite{Schuster2005} to characterize the RX qubit dispersion: we apply a probe tone on resonance with the coupling resonator, drive the qubit via the gate line and tune its energy with $\Delta$. The spectroscopic signal in Fig.~1(e) agrees with the theoretically expected qubit dispersion for qubit configuration 3 (see  Table \ref{QubitConfigsTable}).

\section{Resonant interaction}
First, we investigate the resonant interaction between the coupling resonator and the RX qubit. To start with, both qubits are energetically detuned from the coupling resonator. Then, we sweep $\Delta$ to cross a resonance between the RX qubit and the resonator, while keeping the transmon far detuned. We observe a well resolved avoided crossing in the $|S_{11}|$ reflectance spectrum shown in Fig.~2(a)
\begin{figure}[t!]
\includegraphics[bb=0 0 232 212,width=\linewidth]{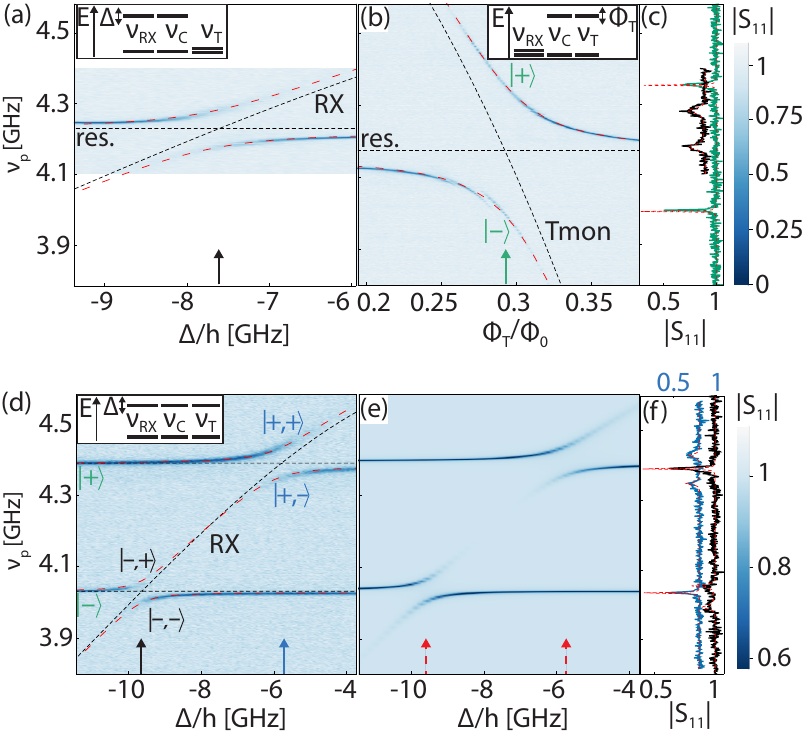}
\caption{\label{Fig2} Resonant interaction. The schematics at the top of the graphs indicate the energy levels of the RX qubit ($\nu_\mathrm{RX}$), coupling resonator ($\nu_\mathrm{C}$) and transmon ($\nu_\mathrm{T}$). Theory curves in the absence (presence) of coupling are shown as dashed black (red) lines. (a) Reflected amplitude $|S_{11}|$ as a function of RX detuning $\Delta$ and probe frequency $\nu_\mathrm{p}$ for RX qubit configuration 2. (b) Reflected amplitude $|S_{11}|$ as a function of relative transmon flux $\Phi_\mathrm{T}/\Phi_0$ and $\nu_\mathrm{p}$. The states $\ket{\pm}$ are discussed in the main text. (c) Cuts from panel (a) at $\Delta/h\simeq-7.6\,\mathrm{GHz}$ (black) and from panel (b) at $\Phi_\mathrm{T}/\Phi_0\simeq0.3$ (green) as marked with arrows in the respective panels. The black trace is offset in $|S_{11}|$ by 0.1. Theory fits are shown as red dashed lines. (d) $|S_{11}|$ as a function of $\Delta$ and $\nu_\mathrm{p}$ for RX qubit configuration 2. The states $\ket{-,\pm}$ and $\ket{+,\pm}$ are explained in the main text. (e) Theory result for parameters as in (d). Values for $|S_{11}|$ are scaled to the experimental data range in (d). (f) Cuts from panel (d) and from panel (e) (without scaling) at $\Delta/h\simeq-9.8\,\mathrm{GHz}$ and $\Delta/h\simeq-5.6\,\mathrm{GHz}$. The experimental cuts are marked with black and blue arrows in (d), the theory cuts are indicated with red arrows in (e). The blue trace is offset in $|S_{11}|$ by 0.2.} 
\end{figure}
and extract a spin qubit-photon coupling strength of $g_\mathrm{RX}/2\pi=52\,\mathrm{MHz}$ from a fit to the vacuum Rabi mode splitting shown in black in Fig.~2(c). The spin qubit and the coupling resonator photons are strongly coupled since $g_\mathrm{RX}>\kappa_\mathrm{C},\gamma_{2,RX}$, where $\gamma_\mathrm{2,RX}/2\pi=11\,\mathrm{MHz}$ is the RX qubit decoherence rate and $\kappa_\mathrm{C}/2\pi=4.6\,\mathrm{MHz}$ is the bare coupling resonator linewidth. The decoherence rate is determined independently with power dependent two-tone spectrosopy. We dispersively detune the coupling resonator with $\Phi_\mathrm{C}$ from the RX qubit and extrapolate the width of the peak observed in the two-one spectroscopy response [c.f. Fig.~1(e)] to zero drive power \cite{Schuster2005}. 

Next, we characterize the interaction between the transmon and the coupling resonator. We tune the transmon through the resonator resonance by sweeping $\Phi_\mathrm{T}$. For this measurement the RX qubit is far detuned in energy. We resolve the hybridized states of the transmon and the resonator photons in the measured $|S_{11}|$ spectrum in Fig.~2(b). They are separated in energy by the vacuum Rabi mode splitting $2 g_\mathrm{T}/2\pi=360\,\mathrm{MHz}$ illustrated in Fig.~2(c) in green. We perform power dependent two-tone spectroscopy to extract the transmon linewidth by probing the read-out resonator. We obtain $\gamma_{2,\mathrm{T}}/2\pi=0.7\,\mathrm{MHz}$, which we estimate to be limited by Purcell decay \cite{Gambetta2006,Sete2014}. Consequently, the strong coupling limit $g_\mathrm{T}>\kappa_\mathrm{C},\gamma_{2,T}$ is also realized for transmon and coupling resonator photons.

\begin{figure*}[!]
\includegraphics[bb=0 0 504 223,width=\linewidth]{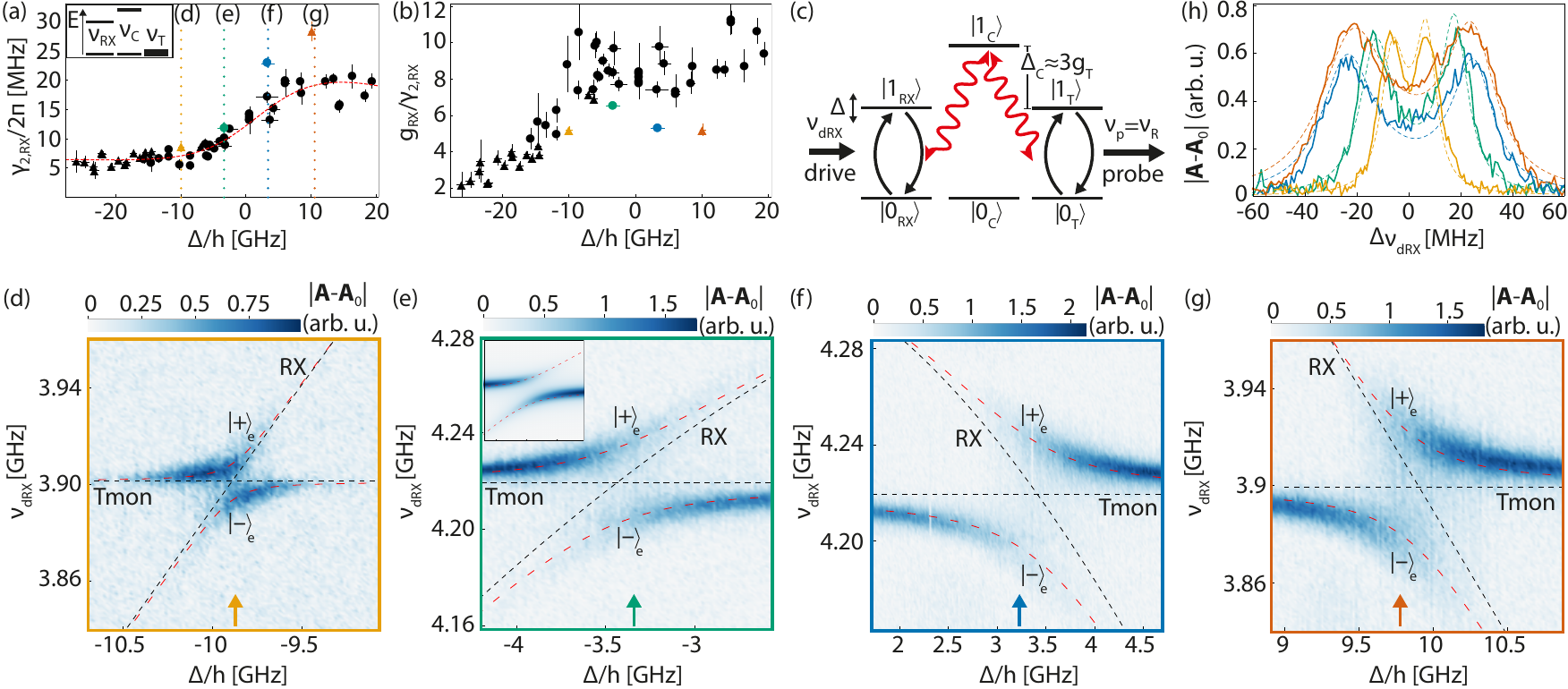}
\caption{\label{Fig3}RX qubit working points and virtual photon-mediated interaction. (a) RX qubit decoherence rate $\gamma_\mathrm{2,RX}$ as a function of detuning $\Delta$. The dotted vertical lines specify the four working points used in (d)-(g). The corresponding colored data points were obtained for a coupling resonator-RX qubit detuning of $\nu_\mathrm{C}-\nu_\mathrm{RX}\simeq(13.7,8.0,5.1,4.4)g_\mathrm{RX}$ for $\Delta/h\simeq(-9.9,-3.3,3.4,10.2)\,\mathrm{GHz}$ and the RX qubit configurations 3 (circle) and 4 (triangle). For the black data points, $\nu_\mathrm{C}-\nu_\mathrm{RX}\geq 9.7\,g_\mathrm{RX}$ with qubit configuration 1 (circle) and 2 (triangle). The dashed red line is a fit of a theory model (see main text) to the black data points. Error bars indicate the standard error of fits and an estimated uncertainty of the RX qubit energy of $50\,\mathrm{MHz}$. (b) Ratio of $g_\mathrm{RX}$, as obtained from theory, and $\gamma_\mathrm{2,RX}$ as shown in (a). The color and shape code of the data points is the same as in (a). (c) Ground ($\ket{0_i}$) and excited states ($\ket{1_i}$) level alignment used in (d)-(g). (d)-(g) Two-tone spectroscopy  at $\nu_\mathrm{p}=\nu_\mathrm{R}\simeq5.6\,\mathrm{GHz}$ as a function of $\Delta$ and drive frequency $\nu_\mathrm{dRX}$. Dashed black (red) lines indicate transmon and RX qubit energies in the absence (presence) of coupling. The frame color refers to the RX qubit working points as specified in (a). The inset in (e) shows the result from theory with the same axes as the main graph. (h) Cuts from panels (d)-(g) at $\Delta$ as specified with arrows in the corresponding panels. The cuts are centered around zero by accounting for a frequency offset $\nu_\mathrm{dRX,0}\equiv\nu_{\mathrm{dRX}}-\Delta\nu_\mathrm{dRX}$. The dashed lines show the corresponding theory results.}
\end{figure*}

We now demonstrate that the two qubits interact coherently via resonant interaction with the coupling resonator. For this purpose, we first
set the transmon and the coupling resonator on resonance, where the hybrid system forms the superposition states $\ket{\pm}=(\ket{0_\mathrm{T},1_\mathrm{C}}\pm\ket{1_\mathrm{T},0_\mathrm{C}})/\sqrt{2}$ of a single excitation in either the resonator or the qubit. Then, we sweep $\Delta$ to tune the RX qubit through a resonance with both the lower energy state $\ket{-}$ and the higher energy state $\ket{+}$. In the $|S_{11}|$ spectrum in Fig.~2(d), avoided crossings are visible at both resonance points. This indicates the coherent interaction of the three quantum systems that form the states $\ket{-,\pm}$ and $\ket{+,\pm}$, where the second label indicates a symmetric or antisymmetric superposition of the RX qubit state with the transmon-resonator $\ket{\pm}$ states. The splitting $2g_\mathrm{\mp}$ between $\ket{-,\pm}$ and $\ket{+,\pm}$ is extracted from the Rabi cuts in Fig.~2(f). We obtain $2g_+/2\pi=84\,\mathrm{MHz}$ at $\Delta/h\simeq-5.6\,\mathrm{GHz}$ and $2g_-/2\pi=63\,\mathrm{MHz}$ at $\Delta/h\simeq-9.8\,\mathrm{GHz}$ from the fits in Fig.~2(f). The smaller $g_-$ compared to $g_+$ is due the decrease of the RX qubit dipole moment with more negative $\Delta$. The experimental observation in Fig.~2(d) is well reproduced by a quantum master equation simulation shown in Fig.~2(e) and further discussed in Ref.~\onlinecite{Supplement}.

\section{RX qubit optimal working point}
While $\gamma_\mathrm{2,T}$ is limited by Purcell decay and therefore does not depend on $\Phi_\mathrm{T}$, $\gamma_\mathrm{2,RX}$ changes with $\Delta$ \cite{Landig2018}. For obtaining the data shown in Fig.~3(a) we use power dependent two-tone spectroscopy via the coupling resonator to measure $\gamma_\mathrm{2,RX}$ as a function of $\Delta$. We observe an increase of $\gamma_\mathrm{2,RX}$ as the charge character of the qubit is increased with $\Delta$. Compared to Ref.~\onlinecite{Landig2018}, the data in Fig.~3(a) covers a larger range in $\Delta$, in particular for $|\Delta|\gg t_\mathrm{l,r}$. The data suggests a lower limit of $\gamma_\mathrm{2,RX}/2\pi\simeq 6.5\,\mathrm{MHz}$ for $\Delta\ll 0$. This is in agreement with Refs.~\onlinecite{Malinowski2017} and \onlinecite{Landig2018}, where the RX qubit was operated at a finite magnetic field of a few hundred mT. Hence, our experiment indicates that the RX qubit can be operated near zero magnetic field without reducing its optimal coherence. In our experiment, the maximum external magnetic field determined by $\Phi_\mathrm{C}$ is of the order of $1\,\mathrm{mT}$. To model the RX qubit decoherence in Fig.~3(a), we consider an ohmic spectral density for the charge noise as well as the hyperfine field of the qubit host material that acts on the spin part of the qubit (see Ref.~\onlinecite{Supplement} for details). Theory and experiment in Fig.~3(a) match for a width $\sigma_B=3.48\,\mathrm{mT}$ of the hyperfine fluctuations in agreement with other work on spin in GaAs \cite{Johnson2005-2,Koppens2005,Koppens2006}. This suggests that 
$\gamma_\mathrm{2,RX}$ is limited by hyperfine interactions.

The colored data points in Fig.~3(a) were measured for a smaller RX qubit-coupling resonator detuning compared to the black data points (numbers are given in Fig.~3 caption). The smaller detuning is used for the virtual interaction measurements explained below. We observe an increase of $\gamma_\mathrm{2,RX}$ for small qubit-resonator detuning compared to large detuning. This increase is about one order of magnitude larger than our estimated difference of Purcell decay and measurement induced dephasing for those different data sets. In contrast, for the transmon that is insensitive to charge noise, we do not observe this effect. This suggests that the effect is due to charge noise induced by the coupling resonator. 

As $\gamma_\mathrm{2,RX}$ increases with $\Delta$ in Fig.~3(a), the RX qubit coupling strength $g_\mathrm{RX}$ to the coupling resonator increases. This implies the possible existence of an optimal working point for the RX qubit, where $g_\mathrm{RX}/\gamma_\mathrm{2,RX}$ is maximal. While a distinct optimal point is not discernible for the black points in Fig.~3(b), the averaged value of $g_\mathrm{RX}/\gamma_\mathrm{2,RX}\simeq 9$ in the spin dominated regime for $-6<\Delta/h<0\,\mathrm{GHz}$ is about a factor of $1.7$ larger than values reported so far for Si spin qubits \cite{Samkharadze2018,Mi2018}. In contrast, for the colored data points we observe an optimal working point at $\Delta/h\simeq -3.3\,\mathrm{GHz}$ since $g_\mathrm{RX}/\gamma_\mathrm{2,RX}$ is reduced at small qubit-resonator detuning in Fig.~3(b) compared to the black data points at large detuning due to the influence of the coupling resonator on $\gamma_\mathrm{2,RX}$ discussed above.

\section{Virtual photon coupling}
In the following we investigate the RX qubit-transmon interaction mediated by virtual photons in the coupling resonator at the RX qubit working points marked in color in Fig.~3(a). The two qubits are resonant while the coupling resonator is energetically detuned, such that the photon excitation plays only a minor role in the superposed two-qubit eigenstates. This coupling scheme, illustrated in Fig.~3(c), is typically used for superconducting qubits to realize two-qubit operations \cite{DiCarlo2009}. We measure the virtual coupling at the optimal working point ($\Delta/h\simeq -3.3\,\mathrm{GHz}$), at $\Delta/h\simeq -9.9\,\mathrm{GHz}$ and at $\Delta/h\simeq 10.2\,\mathrm{GHz}$, where $\gamma_\mathrm{2,RX}$ in Fig.~3(a) saturates, as well as in the intermediate regime at $\Delta/h\simeq 3.4\,\mathrm{GHz}$.  While the RX qubit is tuned through a resonance with the transmon by changing $\Delta$, they are both detuned by $\Delta_\mathrm{C}\equiv \nu_\mathrm{C}-\nu_\mathrm{T}\simeq 3 g_\mathrm{T}$ from the coupling resonator. To realize this detuning for every working point, we adjust the qubit and resonator energies with $\Phi_\mathrm{T}$, $t_\mathrm{l,r}$ and $\Phi_\mathrm{C}$. We drive the RX qubit at frequency $\nu_\mathrm{dRX}$ [see Fig.~1(a)] and investigate its coupling to the transmon by probing the dispersively coupled read-out resonator at its resonance frequency ($\nu_\mathrm{p}=\nu_\mathrm{R}\simeq 5.6\,\mathrm{GHz}$). This measurement is shown in Fig.~3(d) for the working point at $\Delta/h\simeq -9.9\,\mathrm{GHz}$. For large transmon-spin qubit detuning ($\Delta/h\ll-10\,\mathrm{GHz}$), the spectroscopic signal of the transmon is barely visible as the drive mainly excites the bare RX qubit. The signal increases with $\Delta$ as the RX qubit approaches resonance with the transmon, such that driving the RX qubit also excites the transmon due to their increasing mutual hybridization. On resonance, we weakly resolve the two entangled spin qubit-transmon states, which can be approximated as $\ket{\pm}_\mathrm{e}\simeq(\ket{0_\mathrm{RX},1_\mathrm{T}}\pm \ket{1_\mathrm{RX},0_\mathrm{T}})/\sqrt{2}$. These states are separated in energy by the virtual-photon-mediated exchange splitting $2J\simeq 2g_\mathrm{RX}g_\mathrm{T}/\Delta_\mathrm{C}$. The splitting is enhanced at the other working points in Figs.~3(e)-(g), for which the RX qubit control parameter $\Delta$ and consequently $g_\mathrm{RX}$ is larger. The theory inset in Fig.~3(e) agrees well with the experimental observation. The influence of the RX qubit decoherence rate $\gamma_\mathrm{2,RX}$ on the virtual interaction measurement is quantified in Fig.~3(h), where we show averaged cut measurements on resonance, as indicated by arrows in Figs.~3(d)-(g). The theory fits in Fig.~3(h) show excellent quantitative agreement with the experimental curves. As discussed in detail in Ref.~\onlinecite{Supplement}, fit parameters previously obtained from Fig.~2 had to be adjusted to account for significant power broadening in these measurements. The exchange splitting is best resolved at the optimal working point, corresponding to the solid green curve in Fig.~3(h), where we obtain $2J/2\pi\simeq32\,\mathrm{MHz}$ from the fit.

\section{Conclusion}
In conclusion, we have implemented a coherent long-distance link between an RX qubit and a transmon that either utilizes real or virtual microwave photons for the qubit-qubit interaction. The RX qubit was operated in both spin and charge dominated regimes. We found an optimal working point at which the ratio between its resonator coupling and its decoherence rate is maximal and comparable to state of the art values achieved with spin qubits in Si. We also reported that the coupling resonator introduces charge noise that can have significant impact on the RX qubit coherence. The performance of the quantum link in this work is limited by the minimum deoherence rate of the qubit, which is determined by hyperfine interaction in the GaAs host material. Once the spin coherence is enhanced by using hyperfine free material systems such as graphene \cite{Trauzettel2007, Eich2018} or isotopically purified silicon \cite{Zwanenburg2013}, the spin could be used as a memory that can be coupled on-demand to the transmon by pulsing the qubit control parameter. As the coherence of the RX qubit is not altered at zero magnetic field in contrast to other spin qubit implementations, relying on large external magnetic fields, the quantum device architecture used in this work is promising for a high fidelity transmon--spin-qubit and spin-qubit--spin-qubit interface in a future quantum processor.

\begin{acknowledgments}
We acknowledge discussions with Guido Burkard, Michele Collodo, Christian Kraglund Andersen and Maximilian Russ. We also thank David van Woerkom for his contribution to the sample fabrication and for input to the manuscript. This work was supported in part by the Swiss National Science Foundation through the National Center of Competence in Research (NCCR) Quantum Science and Technology. S.N.C. and M.F. acknowledge support of the Vannevar Bush Faculty Fellowship program sponsored by the Basic Research Office of the Assistant Secretary of Defense for Research and Engineering and the funding by the Office of Naval Research through Grant No. N00014-15-1-0029. M.F. and J.C.A.U. acknowledge the support by ARO (W911NF-17-1-0274). The views and conclusions contained in this work are those of the authors and should not be interpreted as necessarily representing the official policies or endorsements, either expressed or implied, of the Army Research Office (ARO) or the U.S. Government. The U.S. Government is authorized to reproduce and distribute reprints for Governmental purposes, notwithstanding any copyright notation thereon.\\
A.J.L., J.V.K. and P.S. fabricated the sample. A.J.L. and J.V.K. performed the measurements with input from B.K. A.J.L. analyzed the data. A.J.L., C.M. and J.C.A.U. wrote the manuscript with input from all authors. C.R. grew the heterostructure under the supervision of W.W. C.M. developed the cQED theory. J.C.A.U. derived the hyperfine noise model under the supervision of S.N.C. and M.F. A.W., T.I. and K.E. supervised the experiment.
\end{acknowledgments}

\section{RX qubit tunnel coupling configurations}
Throughout this work, we use the four RX qubit tunnel coupling configurations listed in Table \ref{QubitConfigsTable}.
\begin{table}[h]
\begin{tabular}{c|c|c}
RX qubit config. & $t_\mathrm{l}/h\,\mathrm{[GHz]}$ & $t_\mathrm{r}/h\,\mathrm{[GHz]}$ \\
\hline
1 & 9.91 & 8.26 \\
2 & 9.22 & 8.73 \\
3 & 8.52 & 8.18 \\
4 & 8.80 & 8.77 \\
\hline
\end{tabular}
\caption{\label{QubitConfigsTable}RX qubit tunnel coupling configurations used in this work.}
\end{table}

\end{document}